\documentstyle[multicol,aps,epsf]{revtex}
\begin{document}
\input psfig.tex
\draft

\title{Nature of Driving Force for Protein Folding -- \\
A Result From Analyzing the Statistical Potential}
\author{Hao Li, Chao Tang, and Ned Wingreen}
\address{NEC Research Institute, 4 Independence Way, Princeton, New
Jersey 08540}

\date{January 2, 1997}
\maketitle

\begin{abstract}
In a statistical approach to protein structure analysis,
Miyazawa and Jernigan (MJ) derived a $20\times 20$ matrix
of inter-residue contact energies between different types
of amino acids.  Using the method of eigenvalue decomposition,
we find that the MJ matrix can be accurately reconstructed
from its first two principal component vectors as 
$M_{ij}=C_0+C_1(q_i+q_j)+C_2 q_i q_j$, with constant $C$'s,
and 20 $q$ values associated with the 20 amino acids. 
This regularity is due to hydrophobic interactions 
and a force of demixing, the latter obeying Hildebrand's 
solubility theory of simple liquids.
\end{abstract}

\pacs{87.15.By, 64.75.+g, 02.10.Sp}

\begin{multicols}{2}

Proteins fold into specific three dimensional  structures to perform
their diverse biological functions.  It is now well established that
for small proteins the information contained in the amino acid sequence
is sufficient to determine the folded structure, which is the structure
with minimum free energy \cite{anf}.  Thus the native structure is
dictated by the physical interactions between amino acids in the
sequence, and understanding the nature of such interactions is crucial
for protein structure prediction.

As a protein contains thousands of atoms and interacts with a huge number
of water molecules, it is not feasible to calculate the free energy
function from first principles.  An often adapted practical approach is
to derive a coarse grained potential (often on the level of amino acids)
using the known structures in the existing protein data banks.  In such
an approach, the energy of a particular substructure in proteins is
derived from the number of its  appearances in the structure data bank
via a Boltzmann factor \cite{rMJ,rMJN,rSipp,rJones}.  A classic example
of such a statistical potential is the Miyazawa-Jernigan (MJ) matrix,
a $20\times 20$ inter-residue contact-energy matrix derived by
Miyazawa and Jernigan \cite{rMJ,rMJN,rMJNote}.  This matrix tabulates
the interaction strength between any two types of amino acids in proteins,
and has been widely applied in protein design  and folding
simulations \cite{rSipp,rfold}.

In this letter, we apply a general method of matrix
analysis, namely, eigenvalue decomposition, to the  MJ
 matrix \cite{rNW}.  The analysis reveals an intrinsic
regularity of the MJ matrix, which yields basic information about
the nature of the  driving force for protein folding. We show that
despite the complicated interactions in proteins,
the major driving force is hydrophobic
interaction and a force of demixing, the latter obeying Hildebrand's
solubility theory of simple liquids \cite{rHilde}.  The result allows us
to attribute the interactions responsible for folding  to quantifiable
properties of individual amino acids. These properties
suggest further experimental tests, and
can be used for analyzing sequence-structure relation.

Eigenvalue decomposition is a general approach to
analyzing  matrices. A given $N\times N$ real symmetric
matrix $M$ can be reconstructed by the following formula
\begin{equation}
M_{ij}=\sum_{\alpha=1}^N\lambda_\alpha V_{\alpha,i}
V_{\alpha,j},
\label{eGen}
\end{equation}
where $M_{ij}$ is the element of the matrix in row $i$ and
column $j$, $\lambda_\alpha$ is the $\alpha$th eigenvalue, and
$ V_{\alpha,i} $ is the $i$th component of the corresponding
eigenvector.  We have analyzed the MJ matrix using eigenvalue
decomposition.  First, we subtract the mean $<M_{ij}>$ from each
element and then analyze the  eigenvalue spectrum of the remaining
matrix.  We find that the eigenvalue spectrum
has two dominant eigenvalues which are  much larger in magnitude
than the rest. Specifically, we find $\lambda_1=-22.49$,
$\lambda_2=18.62$, while the rest of the eigenvalues have absolute
values between $2.17$ and $0.013$.
This suggests (as we shall demonstrate below) that the matrix can be
accurately reconstructed using only the first two eigenvectors,
${\tilde M}_{ij}=<M_{ij}>+\lambda_1V_{1,i}V_{1,j}+
\lambda_2V_{2,i}V_{2,j}$.
Further analysis  shows
that the second eigenvector is related to the first one
by a shift and rescaling, i.e.,
$V_{2,i}=\beta+\gamma V_{1,i}$, with $\beta=-0.30$, $\gamma=-0.90$,
and a correlation coefficient $0.986$. Using this relation, the
expression for ${\tilde M}_{ij}$ can be written simply as
\begin{equation}
{\tilde M}_{ij}=C_0+C_1(q_i+q_j)+C_2q_iq_j,
\label{eFit}
\end{equation}
where $q_i\equiv V_{1,i}$, and the C's are constants, $C_0=-1.492$,
$C_1=5.030$, $C_2=-7.400$.  Thus we can reconstruct the MJ matrix (which
in principle
could have 210 independent elements) by using only twenty
parameters $q_i$, associated with the twenty amino acids,
and three interaction coefficients. Such a simple interaction
form is often the starting point for theoretical
modeling of proteins \cite{rOrland}.

The spectrum of the MJ matrix (two large eigenvalues with
corresponding eigenvectors related to each other) reflects
the specific physical interaction between the amino acids.
The connection between the interaction  and the
spectrum can be understood in the following  general way:
Consider a pairwise interaction
matrix $M_{ij}$ which  is determined by certain properties
of two species i and j, denoted by  $q_i$ and $q_j$.
Assume, on physical grounds, that  $M_{ij}$ can be expressed as
an analytical function $f(q_i,q_j)$
with a well defined converging  power series,
$f(q_i,q_j)=C_0+C_1(q_i+q_j)+C_2q_i q_j+C_3(q_i^2+q_j^2)+
C_4(q_iq_j^2+q_jq_i^2)+...$, where the  $C$'s are constants.
Take first the example  where the expansion ends at the $C_2$
term, i.e., $M_{ij}=C_0+C_1(q_i+q_j)+C_2q_i q_j$.
Since any  row of the
matrix $M$ is given by a vector ${\bf U}_i=(C_0+C_1 q_i){\bf
I}+(C_1+C_2q_i)
{\bf Q}$, which is a linear combination of ${\bf I}$ and
${\bf Q}$, where ${\bf I}\equiv\{1,1,....1\}$,
and ${\bf Q}\equiv\{q_1,q_2....q_n\}$,
one can  decompose the  vector space ${\cal G}$
into the subspace
${\cal G}_{\parallel}$
spanned by ${\bf I}$ and ${\bf Q}$, and its perpendicular
compliment ${\cal G}_\perp$.
It is obvious that ${\cal G}_\perp$ gives rise to
$n-2$ zero eigenvalues, as $M{\bf V}_{\perp}=0$ for
any vector ${\bf V}_{\perp}$ in the subspace
${\cal G}_\perp$. Furthermore, the two  eigenvectors with
nonzero eigenvalues must be expressible as a linear
combination of ${\bf I}$ and ${\bf Q}$, therefore they
are related to each other by a shift and rescaling.
Similarly, if the expansion ends at the $C_4$ term,
there will be three nonzero eigenvalues, and the corresponding
eigenvectors will lie in  a subspace spanned by ${\bf I}$, ${\bf Q}$,
and ${\bf Q}^2$, where  ${\bf Q}^2\equiv\{q_1^2,q_2^2,...,q_n^2\}$.
The same argument applies to all higher order expansions.
This analysis applies to the ideal case where there is no noise
in the matrix.  Introducing noise leads to a slight mixing of
${\cal G}_{\perp}$ and ${\cal G}_{\parallel}$ and therefore to
small nonzero values for the rest of the eigenvalue spectrum.
\begin{figure}
\narrowtext
\vskip -1.5true cm
\centerline{\epsfxsize=4.4in
\epsffile{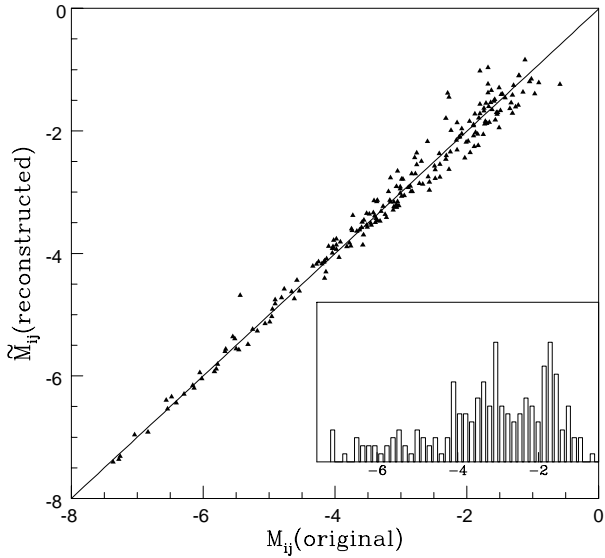}}
\vskip -1.0true cm
\caption{Correlation between $M_{ij}$, the original matrix
elements and ${\tilde M}_{ij}$, the matrix elements
reconstructed from Eq.~(\ref{eFit}).  The regression line is
$y=0.999x-0.008$. The correlation coefficient is $0.989$.
Insert: the distribution of the MJ matrix elements. The unit 
of energy is $k_BT$.}
\end{figure}

The reconstructed matrix in Eq.~(\ref{eFit}) reproduces the
original MJ matrix to a high accuracy.  Fig.~1
shows the correlation between the original MJ matrix
and the reconstructed one. The regression line is
$y=0.999x+0.008$, and the correlation coefficient is $0.989$. On average
Eq.~(\ref{eFit}) gives matrix elements with only $5\%$ error
compared to the original matrix.

Notice that one can redefine the $q$'s in Eq.~(\ref{eFit})
by a shift and rescaling while leaving the interaction
form unchanged.  Therefore any transformation $q\rightarrow A q+B$ with
a corresponding change in the C's yields an identical matrix.  To better
understand the physical meaning of Eq.~(\ref{eFit}), we rewrite it in the
following form,
\begin{equation}
{\tilde M}_{ij}=h_i+h_j-C_2(q_i-q_j)^2/2,
\label{eFnew}
\end{equation}
where
\begin{equation}
h_i=C_0/2+C_1q_i+(C_2/2)q_i^2.
\label{ehh}
\end{equation}
Now each term in Eq.~(\ref{eFnew})
above is invariant under the transformation discussed above.

What is the physical basis for the simple interaction form
in Eq.~(\ref{eFnew})?
Consider the quantity $\chi_{ij}\equiv 2M_{ij}-M_{ii}-M_{jj}$.
Since  $M_{ij}$ is the energy of forming a
contact between type  i and type j amino acids in
water, $\chi_{ij}$ gives the energy of breaking
one  i-i contact and one j-j contact and
forming two pairs of i-j contacts; thus $\chi_{ij}$
is the energy change due to the
mixing of the two types of amino acids. According to Eq.~(\ref{eFnew}),
$\chi_{ij}=-C_2(q_i-q_j)^2$. This form has a striking
similarity to the mixing energy of two simple liquids
as given by  Hildebrand's solubility theory (HST) \cite{rHilde}.
In his 1933 classic paper,
Hildebrand derived the energy of mixing of two simple
liquids by summing over the pairwise interactions
throughout the mixture.  Assuming that the mixing
is random and that the potentials
between  molecules are of the Lennard-Jones type due to the
London dispersion force \cite{rLondon}, Hildebrand arrived
at a formula which expresses the energy of mixing of liquids A and B
as $E_{mixing}\propto (\delta_A-\delta_B)^2$, where $\delta_{A,B}$
are pure component properties related to the
square root of the vaporization energies of  liquid A and B,
traditionally called the ``solubility parameter''.

Now we can imagine the formation of 2 i-j contacts in water by
two steps, formation of an i-i contact and a j-j
contact followed by a mixing. The energy change for the first step is
$2h_i+2h_j$, and that for the second step $\chi_{ij}$. As the formation
of an i-i contact in water is related to the segregation of amino acids
of type i in water, we expect that $h_i$ is related to the hydrophobicity
of amino acid i.  Indeed, we find that $h_i$ correlates very
well with the hydrophobicity scales published in the
literature \cite{rCHY} (see Fig.~2).  Thus despite
the complicated interactions in
proteins, we find that the pairwise inter-residue interactions
responsible for folding can
be attributed to the hydrophobic force and a force of
demixing, the latter obeying HST (Although HST was derived for
simple non-polar molecules, it was found previously
that the theory describes  well the behavior of
polymer blends \cite{rGrassley}.  The
application to proteins is another example of the more
general scope of HST.).
\begin{figure}
\narrowtext
\centerline{\epsfxsize=3.7in
\epsffile{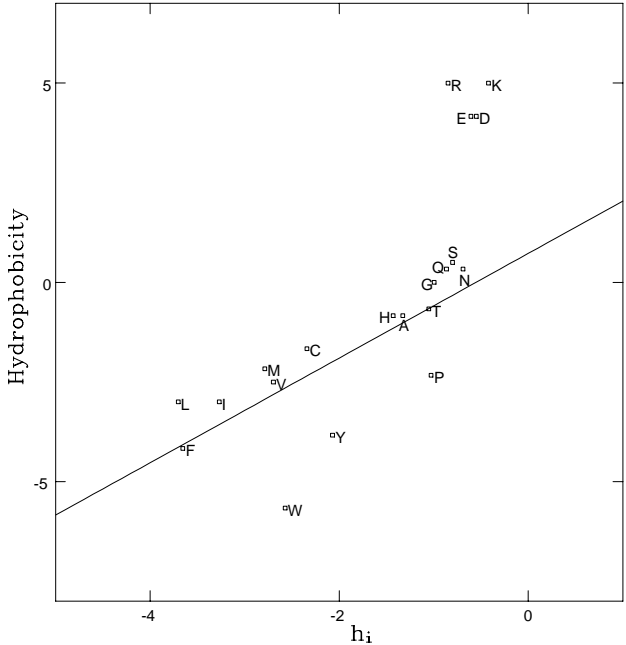}}
\vskip 0.3true cm
\caption{Calculated $h_i$ and measured hydrophobicities [12] of
the 20 amino acids.  The type of amino acid is indicated using the
standard one letter code.  The straight line is a linear fit (excluding
the charged amino acids) with slope $1.314$ and intercept $0.759$. The
correlation coefficient is $0.769$.}
\end{figure}

The above analysis presents a simple picture of the nature of interactions
between amino acids.  It also provides experimentally testable
predictions.  Comparison with HST indicates that the $q_i$
we derive should be linearly related to the solubility parameter
of amino acid $i$, which can be measured. Furthermore,
we predict from Eq.~(\ref{ehh}) that  hydrophobicity can be expressed as
a quadratic function of the solubility parameter.  Since the solubility
parameter and the hydrophobicity of an amino acid can be measured
independently, this prediction can also be tested.

Comparison of the  terms in Eq.~(\ref{eFnew}) shows that
the linear term $h_i+h_j$ is the dominant one in selecting the
native structure. This is because the typical
difference of the linear term $\delta h$ (among different types of
contacts) is much larger than the typical difference of the
square term $\delta \chi/2$, specifically,
 $\delta h =6.52(\delta \chi/2)$. Therefore
the energy difference between different compact structures
(due to different arrangements of the contacts) is mainly
due to the linear term.  Thus, through a  quantitative analysis
of the MJ matrix  we arrive at the conclusion
that the hydrophobic force is the
dominant driving force for protein folding \cite{rDilla}.

The term $-C_2 (q_i-q_j)^2/2$ has an important consequence, however.
This term favors demixing of amino acids ($C_2$ is negative).
The microscopic basis for such a demixing force  is the
dissimilar polarizability of the two monomers \cite{rLondon}.
Since the interior of a protein is composed  of various types of amino acids
which tend to segregate, an amino acid buried in the interior of
a protein will
experience an environment which is quite  different from a uniform
non-polar environment. It  has been controversial
whether one can  model the interior of a protein as a uniform
non-polar environment \cite{rChothia}.  This study suggest that in
general it is not adequate to do so.

Notice that in Fig.~2 the charged amino acids (E, D, R, K)
fall into a distinct group. Since
Eq.~(\ref{eFit}) also gives accurate values for the matrix
elements involving charged amino acids, we believe our $q$ scale captures
more information regarding folding than
a simple hydrophobicity scale.  Other noticeable exceptions are the
cyclic amino acid Proline, and the two amino acids with aromatic
residues Tryptophan and Tyrosine.

The $q$ values we obtain can be used to characterize amino acids.
The distribution of the $q$ values  is bimodal (see Fig.~3),
 which supports the
notion that amino acids naturally fall into two distinct groups:
``polar" (P)  and ``hydrophobic" (H). This division also accounts for
the three different regions in the distribution of
the MJ matrix elements (see the insert to Fig.~1), which reflect  the
three possible combinations of the two groups: polar--polar,
polar--hydrophobic, hydrophobic--hydrophobic. The sharp division
between the two groups as indicated in Fig.~3 also suggests that
amino acids in the same group may play similar roles in structure
determination. There is experimental evidence to this effect
insofar as
certain proteins can be designed by specifying only the HP pattern
of the sequence \cite{rHecht}.  For the purpose of protein design,
the $q$ values can serve as a useful scale for selecting amino acids.
\begin{figure}
\narrowtext
\vskip -1.5true cm
\centerline{\epsfxsize=4.5in
\epsffile{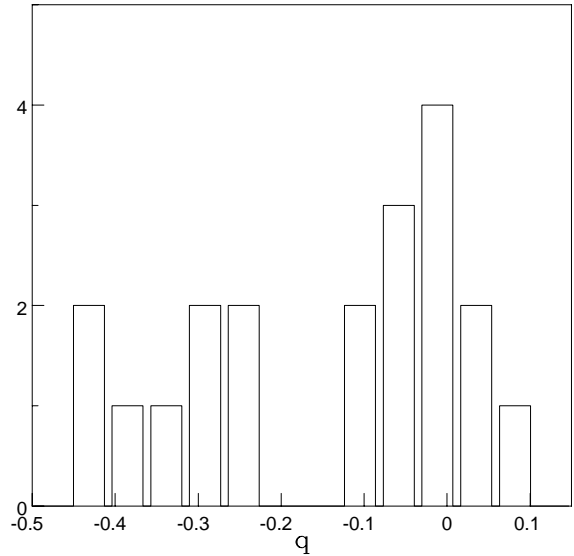}}
\vskip -1.2true cm
\caption{Distribution of $q$ values of the 20 amino acids. The amino 
acids fall into two groups: ``polar'', large $q$, and ``hydrophobic'',
small $q$.}
\end{figure}

The $q$ values can also be used to analyze the relation between sequence
and structure.  In previous studies, hydrophobicity scales have been used
to analyze sequences and locate helical segments \cite{Eisen}.  However,
there exist many different hydrophobicity scales.  Our $q$ scale has the
advantage of being more closely related to the interactions which
determine structure.  We find that for a given sequence, segments with
alternating large and small
$q$ values usually correspond to $\alpha$ helices (consistent with
the previous findings using hydrophobic  scales), segments with
long stretches of large $q$ values usually correspond to reverse turns,
and segments with long stretches of small $q$ values usually
correspond to $\beta$ strands. An example is shown in Fig.~4 of
the 3D structure of the  protein flavodoxin \cite{rLud}, with amino
acids color coded according to
their $q$ values.

To summarize, we were able to extract the regularity of
the Miyazawa-Jernigan matrix of inter-residue contact energies between
amino acids using the method of eigenvalue decomposition.
The analysis reveals that the  driving force for protein folding is
the hydrophobic force and a force of demixing between amino acids.
We were able to construct a solubility
scale for amino acids which can be tested experimentally.
This scale can also be used for selecting amino acids for the purpose
of protein design, and for analyzing sequence-structure relation.
\begin{figure}
\narrowtext
\centerline{\epsfxsize=3.2in
\epsffile{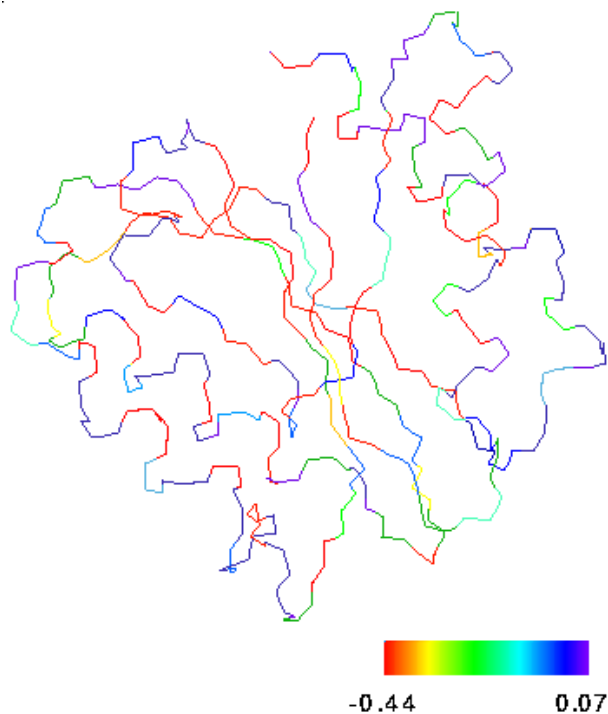}}
\caption{3D structure of the protein 
flavodoxin with only the main chain atoms  plotted.
Amino acids are color coded according to their $q$ values.}
\end{figure}

\end{multicols}


\begin{references}
\bibitem{rMJNote}
To derive the MJ matrix, each residue of a
amino acid in protein structures is represented by the
center of its side chain atom positions.
Two residues are considered to be  in contact if the
distance between corresponding center points is smaller
than $r_C\approx 6.5$\AA. The contact energies $E_{ij}$
between type i and type j amino acids were obtained
by counting the number of contacts between two different
types of amino acids in the structure data bank.
\bibitem{rNW}
In reference \cite{rMJN}, Miyazawa and Jernigan updated their matrix 
published in reference \cite{rMJ} with a significantly larger set
of protein crystal structures. We use the new matrix for
the analysis in this paper. A similar analysis using the
old matrix gives nearly identical results.
\end{references}
\end{document}